\documentclass[a4paper,11pt]{article}
\usepackage{longtable}
\pdfoutput=1 

\usepackage{jcappub} 

\usepackage[T1]{fontenc} 

\title{\boldmath Estimating the baryon fraction in the IGM from well-localized FRBs and DESI data}

\author[a,1]{Thais Lemos,\note{Corresponding author}}

\affiliation[a]{Observat\'orio Nacional, Rio de Janeiro - RJ, 20921-400, Brazil}

\emailAdd{thaislemos@on.br}

\abstract{Current measurements of Baryon Acoustic Oscillations (BAO) from the Dark Energy Spectroscopic Survey (DESI DR2), when combined with data from Type Ia supernovae (SNe), challenge the observational viability of the $\Lambda$-Cold Dark Matter ($\Lambda$CDM) model, motivating combinations of independent datasets to estimate cosmological quantities. In a previous communication, we presented a cosmological independent method to constrain the baryon fraction in the IGM ($f_{\mathrm{IGM}}$), where we derived relevant expressions for the dispersion measure ($\mathrm{DM}$) in terms of luminosity distance, allowing us to estimate $f_{\mathrm{IGM}}$ combining directly measurements of 17 well-localized FRBs and 1048 SNe from the Pantheon catalog. Here we revisit this method to constrain $f_{\mathrm{IGM}}$, considering two parameterizations for the $f_{\mathrm{IGM}}$: constant and time-dependent. We expand our sample by combining 107 well-localized Fast Radio Bursts (FRBs) with BAO measurements from DESI DR2 and SNe observations from DESY5, and the Pantheon+ catalog. We find through a Bayesian model selection analysis that a conclusive answer about the evolution of $f_{\mathrm{IGM}}$ cannot be achieved from the current FRBs observational data. In particular, our results show weak evidence in favor of the constant case.}

\begin{document}
\maketitle
\flushbottom

\section{Introduction}
\label{sec:intro}

The Baryon Acoustic Oscillations (BAO) measurements from the Dark Energy Spectroscopic Instrument (DESI) \cite{DESI_instrument}, combined with data from the cosmic microwave background (CMB) and Type Ia supernovae (SNe), provide important results that challenge the $\Lambda$CDM paradigm. More precisely, the DESI collaboration reported strong evidence for an evolving dark energy component showing tighter constraints on the dark energy equation of state (EoS) using different $w(a)$ parameterizations~\cite{Chevallier:2000qy,Linder:2002et,Barboza:2008rh}. In addition to the discrepancy with $\Lambda$CDM, DESI Data Release 2 (DESI DR2) \cite{DESI} provides the most precise and up-to-date BAO data set. In this context, this discrepancy in the current cosmological paradigm motivates the combination of independent cosmological datasets as well as the use of cosmological model-independent approaches as a precise and accurate method to constrain the fundamental cosmological parameters. 

We combine DESI measurements with fast radio burst (FRB) data to constrain cosmological/astrophysical parameters. The FRBs are radio transient events with millisecond durations and an unknown origin (for a review, see \cite{Petroff2022,Thornton2013,Petroff2015,Petroff2016,Platts2019}). However, the large observed dispersion measure ($\mathrm{DM}$) comparable to the Milky Way contribution suggests a cosmological origin for these events \cite{Dolag2015}. Although $\sim 1000$ FRBs have been detected since the discovery in 2007 by the Parkes telescope \cite{Lorimer2007}, only a few of them in the literature are well localized (with redshift). The $\mathrm{DM}$ combined with the redshift of the host galaxy can be used as an astrophysical and cosmological probe
(see \cite{Walters2018,Wei2018,Lin2021,Wu2021,Reischke2023,Lemos2023,Lemos2023_2} for applications of FRBs in the cosmology). In practice, some issues hinder the application of FRBs for cosmological purposes. For instance, the poor knowledge about the dispersion measure's variance related to the spatial variation in the cosmic electron distribution. To circumvent this problem, such fluctuations ($\delta$) can be parameterized as a function of redshift \cite{Takahashi2021}. Another restriction is the variation with respect to the redshift of the fraction of baryon mass in the intergalactic medium (IGM), called as $f_{\mathrm{IGM}}$, which is degenerated with the cosmological parameters. In \cite{Shull2012}, the authors found  $f_{\mathrm{IGM}} \approx 0.82$ at $z \geq 0.4 $, while in \cite{Meiksin2009} the authors estimated $f_{\mathrm{IGM}} \approx 0.9$ at $z \geq 1.5$.

In a previous communication \cite{Lemos2023}, we proposed a cosmology-independent method that combines directly the dispersion measure of 17 FRB observations and the luminosity distance ($D_{L}$) from supernovae type Ia (SNe) measurements of the Pantheon catalog to constrain a possible evolution of the baryon fraction in the IGM. In the present paper, we revise such method extending our dataset to study the impact of different $D_{L}$ measurements in this type of analysis, combining 107 well-localized FRBs with BAO measurements from DESI DR2 \cite{DESI}, and with SNe catalogs from Pantheon+ \cite{Pantheon+}, and Dark Energy Survey Supernova 5-Year (DES Y5) \cite{DES}. We organized this paper as follows. In Section \ref{sec:theory} we briefly present the FRB's concepts. The datasets used and our model-independent method are presented in Section \ref{sec:data}. In Section \ref{sec:results}, we present our main results. Finally, we summarize the main conclusions in Sec. \ref{sec:conclusions}.  

\section{FRB's theory}
\label{sec:theory}

The FRB's photons interact with the free electrons in the medium along the path from the source to the observer on Earth. These interactions lead to a change in the frequency of the photons, causing a delay in their arrival time. This time delay is related to the $\mathrm{DM}$, for which the observed dispersion measure can be expressed in terms of other components \cite{Deng2014,Gao2014}

\begin{equation}\label{eq:dm}
    \mathrm{DM}_{\mathrm{obs}} = \sum_{i} \mathrm{DM}_{i} (z),
\end{equation}
where the index $i = $ ISM, halo, IGM, and host represents the Milky Way interstellar medium (ISM), the Milky Way halo, the intergalactic medium, and host galaxy contributions, respectively. 

The quantity $\mathrm{DM}_{\mathrm{obs}}(z)$ of a FRB is directly measured from the corresponding event, while $\mathrm{DM}_{\mathrm{MW,ISM}}$ can be well-determined using models of the ISM galactic electron distribution in the Milky Way from pulsar observations \cite{Taylor1993,Cordes2002,Yao2017}. Moreover, the Milky Way halo contribution is not entirely understood yet, and we assume $\mathrm{DM}_{\mathrm{halo}} = 50$ pc/cm$^{3}$ \cite{Macquart2020}.

The host galaxy contribution is not a well-understood parameter due to the challenges in measurement and modeling, because it depends on the type of galaxy, the relative orientations of the FRB source with respect to the host and source, and the near-source plasma \cite{Xu2015}. For this reason, we consider

\begin{equation}
    \mathrm{DM}_{\mathrm{host}} (z) = \frac{\mathrm{DM}_{\mathrm{host,0}}}{1+z},
\end{equation}
where the term $(1+z)$ accounts for the cosmic dilation \cite{Deng2014,Ioka2003} and $\mathrm{DM}_{\mathrm{host,0}}$ is the host galaxy contribution in the source frame and will be a free parameter in our analysis.

The last and largest contribution of the dispersion measure is the IGM, where the cosmological contribution appears. The average of IGM dispersion measure can be written as \cite{Deng2014}

\begin{equation}\label{eq:igm}
\mathrm{DM}_{\mathrm{IGM}}(z) = \frac{3c\Omega_{b}H_{0}^{2}}{8\pi Gm_{p}} \int_{0}^{z} \frac{(1+z')f_{\mathrm{IGM}}(z')\chi(z')}{H(z')}  dz'\;,
\end{equation}
where $c$,  $\Omega_{b}$, $H_{0}$, $G$, $m_{p}$, $f_{IGM}(z)$, $H(z)$ are the speed of light, the present-day baryon density parameter, the Hubble constant, the gravitational constant, the proton mass, the baryon fraction in the IGM and the Hubble parameter at redshift $z$, respectively. While $\chi(z) = Y_{H}\chi_{e,H}(z) + Y_{He}\chi_{e,He}(z)$ is the free electron number fraction per baryon, in which $Y_{H} = 3/4$ and $Y_{He} = 1/4$ are the mass fractions of hydrogen and helium, respectively, and $\chi_{e,H}(z)$ and $\chi_{e,He}(z)$ are the ionization fractions of hydrogen and helium, respectively. The hydrogen and helium are fully ionized at $z < 3$ \cite{Meiksin2009,Becker2011}, so that we have $\chi_{e,H}(z) = \chi_{e,He}(z) = 1$. 

From Eq. \ref{eq:dm}, we can define the observed extragalactic dispersion measure to perform a statistical comparison between the observational data and the theoretical parameters:

\begin{equation}
\label{eq:ext_obs}
  \mathrm{DM}^{\mathrm{obs}}_{\mathrm{ext}}(z) \equiv  \mathrm{DM}_{\mathrm{obs}}(z) - \mathrm{DM}_{\mathrm{ISM}}  - \mathrm{DM}_{\mathrm{halo}} \; ,
\end{equation}
and the theoretical extragalactic dispersion measure 
\begin{equation}
\label{eq:ext_th}
    \mathrm{DM}_{\mathrm{ext}}^{\mathrm{th}}(z) \equiv \mathrm{DM}_{\mathrm{host}}(z) + \mathrm{DM}_{\mathrm{IGM}}(z) \;.
\end{equation}
\section{Data and Methodology}
\label{sec:data}

\subsection{Data}
\indent

In what follows, we will describe the observational datasets for dispersion measure and luminosity distance that we use in this work. 

\subsubsection{FRBs}

The currently available sample of FRBs contains 116 well-localized events (for details of FRBs catalog \footnote{https://blinkverse.alkaidos.cn}, see \cite{Blinkverse}). We exclude from our analysis the following events: FRB 20171020A \cite{FRB171020A}, FRB 20181030 \cite{Bhardwaj2021_2} and FRB \cite{FRB210807D} have a low redshift ($z = 0.0087$, $z = 0.0039$ and $z = 0.023686$, respectively) and can not be associated with any SNe Ia in the DES catalog; FRB 20190520B \cite{FRB190520B} has a host galaxy contribution significantly larger than the other events; FRB 20190614 \cite{FRB190614} has no measurement of spectroscopic redshift and can be associated with two host galaxies; FRB 20200120E \cite{Bhardwaj2021} is estimated in direction of M81, but a Milky Way halo can not be rejected; and finally, FRB 20210405I \cite{FRB210405I} and FRB 20220319D \cite{FRB220207C} have the MW contribution larger than observed one.

We list in Table \ref{tab:data} our working sample that contains 107 FRBs \cite{FRB121102,FRB20191228,FRB180814,FRB180916,FRB180924,FRB181112,FRB181220A,FRB190102,FRB190110C,FRB190523_1,FRB190523_2,FRB190608,FRB201123A,FRB201124,FRB210117A,FRB210320,FRB210410D,FRB210603A,FRB210807D,FRB211203C,FRB220204A,FRB220207C,FRB20220610A,FRB220717A,FRB220912A,FRB221219A,FRB230203A,FRB240114A,FRB240209A} with their main properties: redshift, the Milky Way galaxy contribution ($\mathrm{DM}_{\mathrm{MW,ISM}}$) estimated from the NE2001 model \cite{Cordes2002}, observed dispersion measure ($\mathrm{DM}_{\mathrm{obs}}$), $\mathrm{DM}_{\mathrm{obs}}$ uncertainty ($\sigma_{\mathrm{obs}}$) and the references. For the 41 FRBs marked with the symbol $\dag$ in $\sigma_{\mathrm{obs}}$, the uncertainty in $\mathrm{DM}{\mathrm{obs}}$ is not provided. In these cases, the uncertainty is randomly drawn from a Gaussian distribution constructed using the mean and standard deviation of the $\sigma_{\mathrm{obs}}$ values from the other events.

The observational quantity $\mathrm{DM}_{\mathrm{ext}}$ (Eq. \ref{eq:ext_obs}) can be calculated from Table \ref{tab:data} and its total uncertainty can be expressed by the relation

\begin{equation}\label{uncertainty}
    \sigma_{\mathrm{tot}}^{2} = \sigma_{\mathrm{obs}}^{2} + \sigma_{\mathrm{MW}}^{2} + \sigma_{\mathrm{IGM}}^{2} + \bigg( \frac{\sigma_{\mathrm{host},0}}{1+z} \bigg)^{2} + \delta^{2} \;,
\end{equation}
where $\sigma_{\mathrm{MW}}$ is the average galactic uncertainty and assumed to be 30 pc/cm$^{3}$ \cite{Manchester2005}, $\delta$ is the $\mathrm{DM}$ fluctuations related to the spatial variation in cosmic electron density along the line-of-sight. As long as such fluctuations are not well-constrained by observations, we will treat them as a fixed value, $\delta = 230\sqrt{z}$ pc/cm$^{3}$ \cite{Takahashi2021,Lemos2023}. The uncertainty of $\mathrm{DM}_{\mathrm{host},0}$ can be assumed as $\sigma_{\mathrm{host},0} = 30$ pc/cm$^{3}$ \cite{Li2019}. Finally, the uncertainty of IGM contribution ($\sigma_{\mathrm{IGM}}$) can be calculated from error propagation of Eqs. \ref{eq:igm_var} and \ref{eq:igm_cst} given by

\begin{eqnarray}
    \sigma_{\mathrm{IGM}} = A f_{\mathrm{IGM},0} \bigg[  \frac{\sigma_{D_{L}}^{2}}{c^{2}} + 
    \frac{\sigma_{\mathrm{I}}^{2}}{c^{2}}   \bigg]^{1/2},
\end{eqnarray}
where $\sigma_{D_{L}}$ is the luminosity distance uncertainty that is calculated from the SNe, $\sigma_{\mathrm{I}}$ is the uncertainty of the quantity in Eq. \ref{eq:sum}. 

\subsubsection{BAO}

For BAO dataset we use the results from DESI DR2 \cite{DESI} that contains measurements from more than 14 million galaxies and quasars and are expressed in terms of three distances: the three-dimensional BAO mode ($D_{V}/r_{d}$), the radial mode ($D_{H}/r_{d}$) and the transverse mode ($D_{M}/r_{d}$). In the present work, we only consider the transverse mode due to its relation with the luminosity distance. From BAO scale measurements in the transverse direction at redshift $z$, the transverse comoving distance can be written assuming a flat Universe as

\begin{equation}
    D_{M}  (z) = c \int_{0}^{z} \frac{dz'}{H(z')},
\end{equation}
where it can be related to the luminosity distance from the below expression

\begin{equation}\label{eq:dl_dm}
   D_{L} (z) = (1+z) D_{M} (z) .
\end{equation}

\subsubsection{SNe Ia}

Regarding SNe Ia observations, we use two different catalogs: Pantheon+ \cite{Pantheon+} (the complete catalog is available in the GitHub repository\footnote{https://github.com/PantheonPlusSH0ES/DataRelease}), which contains the analysis of 1701 light curves of 1550 distinct SNe within the redshift range $0.001 < z < 2.3$.; and DES Y5 \cite{DES} (for the complete catalog, see the  GitHub repository\footnote{https://github.com/des-science/DES-SN5YR}), which comprises 1635 events, spanning a redshift range $0.02 < z < 1.13$, along with 194 high-quality external supernovae at redshifts below 0.1, for
a total of 1829 supernovae. In the case of DES Y5 dataset, we remove the SNe with larger uncertainty ($\sigma_{\mu} > 2$ mag), and the remaining sample contains 1757 SNe.
 
From SNe dataset, we can obtain $D_{L}$ by the distance moduli ($\mu(z)$) relation, 

\begin{equation} \label{eq:mz}
    \mu(z) = m_{B} - M_{B} = 5\log_{10}\left[ \frac{D_{L}(z)}{1\mbox{Mpc}}\right] + 25 \;,
\end{equation}
where $m_{B}$ is the apparent magnitude and $M_{B}$ ais absolute magnitude, which here we fix consistent to $H_{0}$ from SH0ES collaboration \cite{Riess2022}, $M_{B} = -19.253 \pm 0.027$ mag. 

The luminosity distance uncertainty relation can be expressed as: 

\begin{equation}
    \sigma_{D_{L}} = \frac{\ln{10}}{5}D_{L} \cdot \sqrt{\sigma_{m_{B}}^{2} + \sigma_{M_{B}}^{2}},
\end{equation}
being $\sigma_{m_{B}}$ and $\sigma_{M_{B}}$ the apparent and absolute magnitude uncertainties, respectively.

 \subsection{Methodology}

In Reference \cite{Lemos2023}, we proposed a cosmological model-independent approach, which we solve $\mathrm{DM}_{\mathrm{IGM}}$ integral by parts, rewriting the $H(z)$ in terms of luminosity distance ($D_{L}$)\footnote{The same method can be done, but instead of luminosity distance definition one can use the transverse comoving distance ($D_{M}$) from Eq. \ref{eq:dl_dm}.} assuming two parameterizations for the baryon fraction: time-dependent and constant cases. In the present paper, we consider the same approach, for which Eq. \ref{eq:igm} can be written for the time-dependent case, $f_{\mathrm{IGM}} (z)= f_{\mathrm{IGM,0}} + \alpha \frac{z}{1+z}$, as 

\begin{equation}\label{eq:igm_var}
     \mathrm{DM}_{\mathrm{IGM}}(z)= A \left[  \left( f_{\mathrm{IGM},0} + \alpha \frac{z}{1+z} \right) \frac{D_{L}(z)}{c} 
    - \left( f_{\mathrm{IGM},0} + \alpha  \right)\int_{0}^{z} \frac{D_{L}(z')}{(1+z')c} dz' \right],
\end{equation}
where $\alpha$ is parameter that quantifies a possible evolution of $f_{\mathrm{IGM}}$. In our analysis, both are free parameters and since $f_{\mathrm{IGM}}$ is understood to be an increasing function of the redshift, $\alpha$ assumes only positive values ($\alpha \geq 0$). 

In the constant case ($f_{\mathrm{IGM}} (z) = f_{\mathrm{IGM,0}}$), Eq. \ref{eq:igm} can be written as

\begin{equation}\label{eq:igm_cst}
    \mathrm{DM}_{\mathrm{IGM}}(z)= A f_{\mathrm{IGM},0} \left[\frac{D_{L}(z)}{c} - \int_{0}^{z} \frac{D_{L}(z')}{(1+z')c} dz' \right].
\end{equation}

The second term in both expressions above can be numerically solved as (see \cite{Holanda2013}):
\begin{equation}\label{eq:sum}
    \int_{0}^{z} \frac{D_{L}(z')}{(1+z')} dz' =  \frac{1}{2}\sum_{i=1}^{N} \left( z_{i+1}-z_{i}\right)\nonumber \times \left[\frac{D_{L}(z_{i+1})}{(1+z_{i+1})} 
+ \frac{D_{L}(z_{i})}{(1+z_{i})}  \right]\;.
\end{equation}

From the above expressions, one can constrain the baryon fraction in the IGM by modeling the terms in $\mathrm{DM}_{\mathrm{ext}}^{\mathrm{th}}$ quantity (Eq. \ref{eq:ext_th}) and comparing these theoretical predictions with the observed values of $\mathrm{DM}_{\mathrm{ext}}$. By combining well-localized FRBs with SNe and BAO (using Eq. \ref{eq:dl_dm} for the last one) datasets, we can obtain constraints on $f_{\mathrm{IGM}}$ and $\mathrm{DM}_{\mathrm{host,0}}$ in a cosmology model-independent mode from the observational data sets described previously. 

In order to estimate $D_{L}$, $D_{M}$ and its uncertainty at the same redshift of FRBs, we perform a Gaussian Process (GP) reconstruction of the Pantheon+, DES Y5, and DESI DR2 datasets, respectively, using GaPP python library\footnote{For details of GaPP (https://github.com/astrobengaly/GaPP), see \cite{GaPP}.}. We perform a Monte Carlo Markov Chain (MCMC) analysis using the \textit{emcee} package \cite{Foreman-Mackey2013} to constrain the free parameters in our analysis: $f_{\mathrm{IGM,0}}$ and $\mathrm{DM}_{\mathrm{host,0}}$ (for Constant case), and $f_{\mathrm{IGM,0}}$, $\alpha$ and $\mathrm{DM}_{\mathrm{host,0}}$ (for Time-dependent case).

Since we are interested in a model-independent approach and to be consistent with our choice of $\mathrm{M}_{\mathrm{B}}$, we adopt the value for the Hubble constant in Eqs. \ref{eq:igm_var} and \ref{eq:igm_cst} from SH0ES collaboration, $H_{0} = 73.04 \pm 1.01$ km/s/Mpc \cite{Riess2022}. We also assume model-independent values for the the sound horizon at the drag epoch, $r_{d} = 101 \pm 2.3h^{-1}$ Mpc \cite{Verde2017}, to reconstruct $D_{M}$ and for the baryon density parameter, $\Omega_{b}h^{2} =  0.02235 \pm 0.00037$, from Big Bang Nucleosynthesis (BBN) analysis reported by \cite{Cooke2018}.

\section{Results}
\label{sec:results}

\begin{figure*}
\begin{center}
\includegraphics[width=0.48\textwidth]{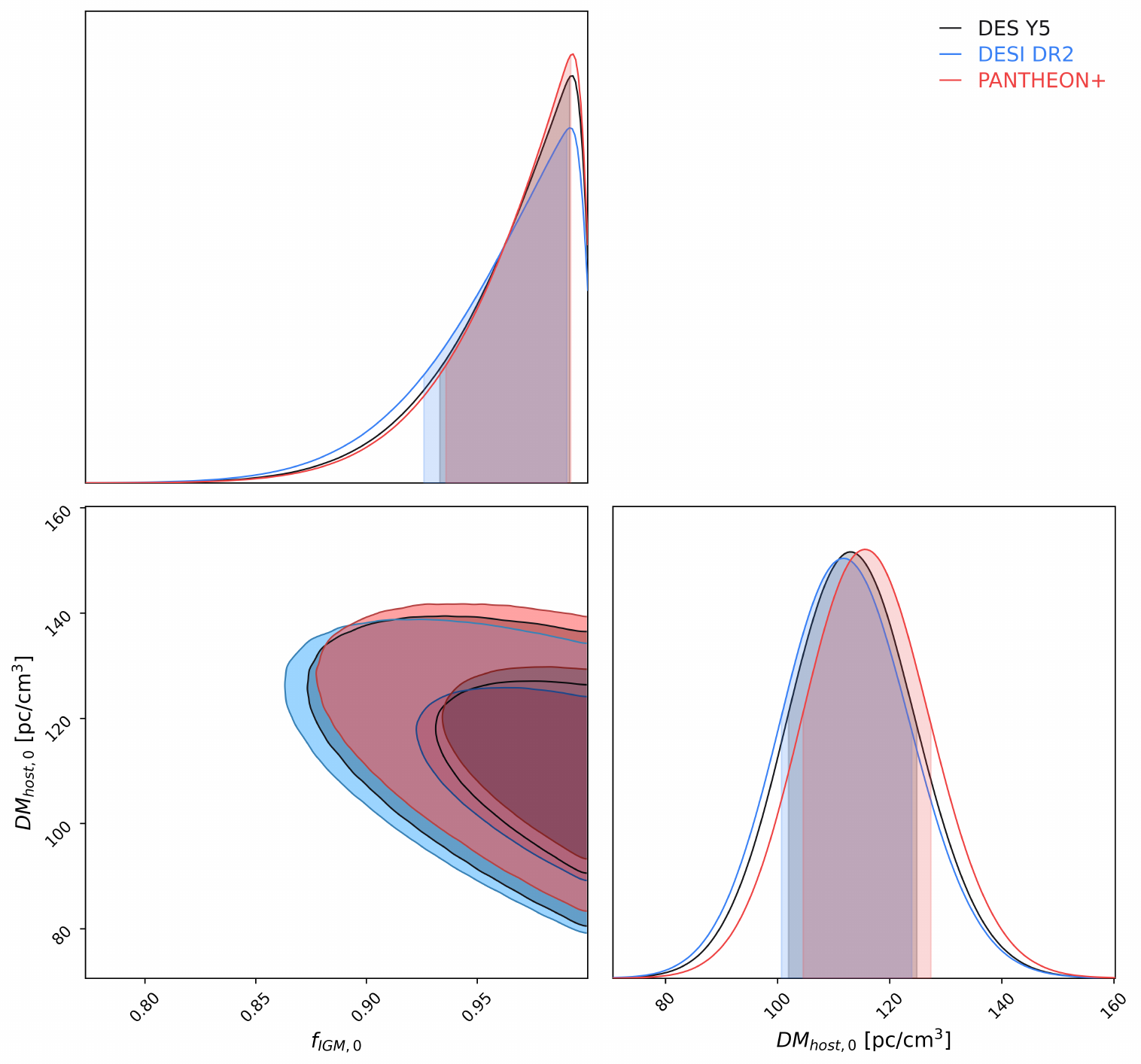}
\includegraphics[width=0.48\textwidth]{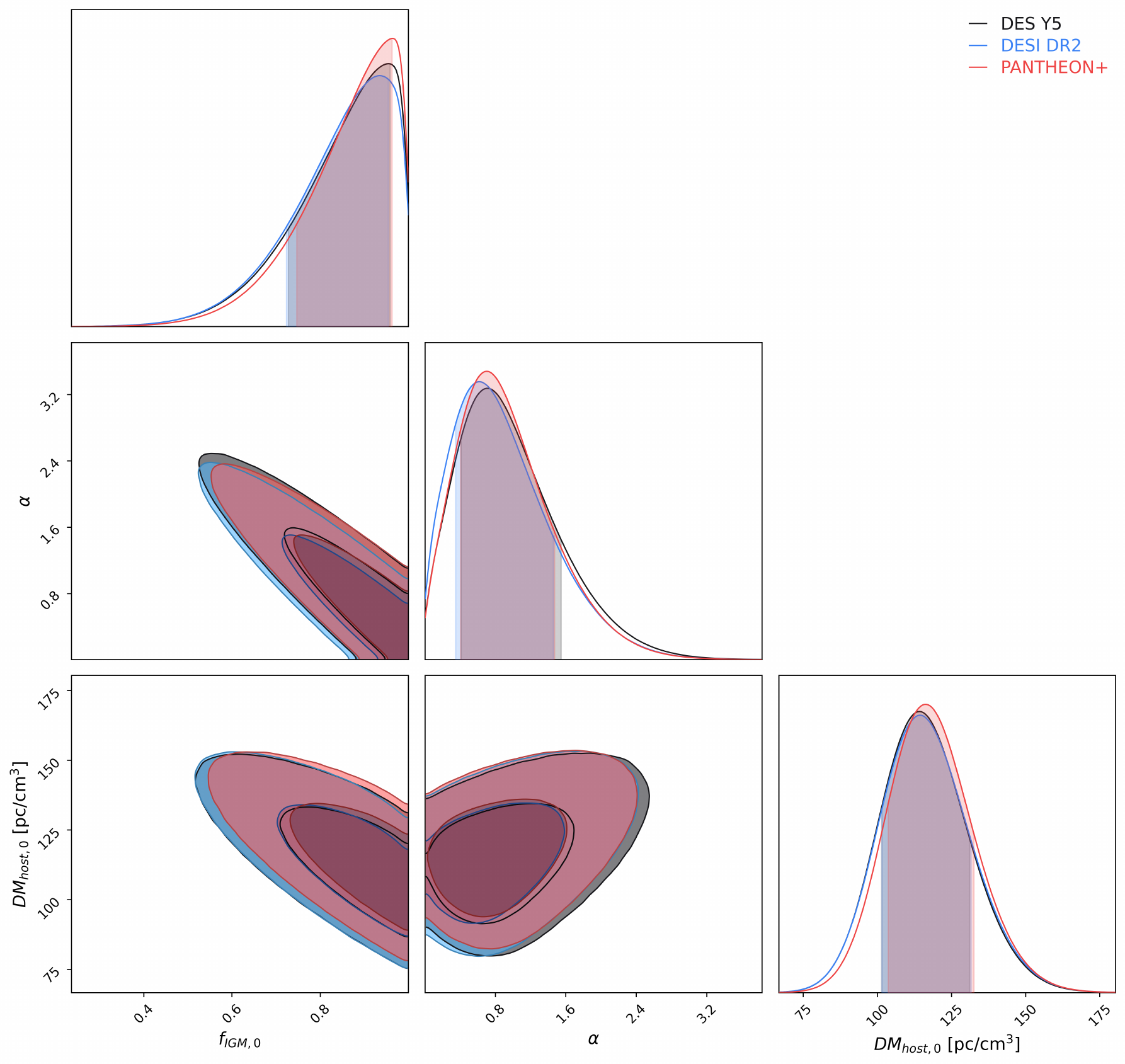}
\caption{Constraints on the baryon fraction $f_{\mathrm{IGM}}$ parameters and the mean host galaxy contribution of dispersion measure $\mathrm{DM}_{\mathrm{host,0}}$ for the three data combination. {\it{Left:}} considering the constant case. {\it{Right:}} for the time-dependent parameterization.} 
\label{fig:results}
\end{center}
\vspace{0.3cm}
\end{figure*}


\begin{table*}[h]
\vspace{0.3cm}
\centering
\caption{Estimates of the  $f_{\mathrm{IGM}}$ parameters and $\mathrm{DM}_{\mathrm{host,0}}$.}
\begin{tabular}{ c  c  c  c  }
\hline
Dataset & $f_{IGM,0}$ & $\alpha$ & $DM_{host,0}$ [pc/cm$^{3}$] \\
\hline
\textbf{Constant case}   &  &  &   \\
FRB + DES Y5  & $0.999_{-0.066}^{+ 0.008}$ & - & $113_{-11}^{+ 12}$ \\
FRB + DESI DR2  & $0.999_{-0.073}^{+0.010}$ & - & $112_{-11}^{+ 12}$ \\
FRB + Pantheon+ & $0.999_{-0.066}^{+ 0.007}$ &  - & $116_{-11}^{+ 12}$ \\
\hline
\hline
\textbf{Time-dependent case}   &  &  &  \\
FRB + DES Y5  & $0.955_{-0.227}^{+ 0.003}$ & $0.70_{-0.30}^{+0.84}$ & $114_{-13}^{+ 17}$ \\
FRB + DESI DR2  & $0.935_{-0.211}^{+0.020}$ & $0.61_{- 0.26}^{+ 0.85}$& $114_{-13}^{+ 18}$ \\
FRB + Pantheon+ & $0.972_{-0.225}^{+ 0.009}$ & $0.71_{-0.31}^{+0.76}$ & $116_{-12}^{+ 16}$ \\
\hline
\end{tabular}
\label{tab:results}
\end{table*}

Figure \ref{fig:results} shows the posterior probability density function and $1-2\sigma$ contours for combinations of the free parameters $f_{\mathrm{IGM,0}}$, $\alpha$ and $\mathrm{DM}_{\mathrm{host,0}}$ considering parameterizations constant (left Panel) and time-dependent (right Panel). In Table \ref{tab:results} we present the results of the free parameters for both cases, considering the different datasets. For the constant case, the best-fit for the baryon fraction and host galaxy contribution is almost the same for the three datasets ($f_{\mathrm{IGM,0}} \sim 0.99^{+0.01}_{-0.07}$ and $\mathrm{DM}_{\mathrm{host,0}} \sim 113 \pm 12$ pc/cm$^{3}$ at $1\sigma$ level). For the time-dependent case, we estimate $f_{\mathrm{IGM,0}}$ ranging from $0.935_{-0.211}^{+0.020}$ (FRB + DESI DR2) to $0.972_{-0.225}^{+ 0.009}$ (FRB + Pantheon+) and from $0.61_{- 0.26}^{+ 0.85}$ (FRB + DESI DR2) to $0.71_{-0.31}^{+0.76}$ (FRB + Pantheon+) for $\alpha$ at $1\sigma$. While the host galaxy contribution we estimated ranged from $114_{-13}^{+17}$ pc/cm$^{3}$ (FRB + DES Y5) to $116_{-12}^{+16}$ pc/cm$^{3}$ (FRB + Pantheon+) ($1\sigma$). For both  constant and time-dependent cases, the constraints on $f_{\mathrm{IGM,0}}$ and $\mathrm{DM}_{\mathrm{host,0}}$ are compatible in $2\sigma$ level. 

Comparing the results for the constant case with those obtained by Ref. \cite{Lemos2023}, the values only agree at $2\sigma$ level for the cases where the fluctuation is $\delta = 50$ and $100$ pc/cm$^{3}$ in Ref. \cite{Lemos2023}, which is expected since here we are assuming $\delta = 230\sqrt{z}$ pc/cm$^{3}$. Now, considering the time-dependent case, in Ref. \cite{Lemos2023} we obtained for $f_{\mathrm{IGM,0}}$ varies from $f_{\mathrm{IGM,0}} = 0.483 \pm 0.066$ ($\delta = 0$ pc/cm$^{3}$) to $f_{\mathrm{IGM,0}} = 0.40 \pm 0.26$ ($\delta = 100$ pc/cm$^{3}$), from $\alpha = 1.21 \pm 0.28$ ($\delta = 0$ pc/cm$^{3}$) to $\alpha = 1.56 \pm 1.08 $ ($\delta = 100$ pc/cm$^{3}$) and from $\mathrm{DM}_{\mathrm{host,0}} = 190.1 \pm 9.1$ pc/cm$^{3}$ ($\delta = 0$ pc/cm$^{3}$) to $\mathrm{DM}_{\mathrm{host,0}} = 196.56 \pm 53.62$ pc/cm$^{3}$ ($\delta = 100$ pc/cm$^{3}$) at $1\sigma$ level. Comparing these values with the results estimated in the present paper, we can note that the results are not consistent at  $2\sigma$ since the present value for $f_{\mathrm{IGM,0}}$ increased, being now in agreement with the values for the constant case (Tab. \ref{tab:results}) and with other studies that used the same parameterization (see e.g. \cite{Wei2019,Dai2021} at $2\sigma$ level. While here we estimate values of the parameter $\alpha$ lower than the values we obtained in Ref. \cite{Lemos2023}, agreeing with the values in Ref. \cite{Wei2019,Dai2021}  ($2\sigma$). Note that the errors on the $f_{\mathrm{IGM,0}}$, $\alpha$, and $\mathrm{DM}_{\mathrm{host,0}}$ parameters depend on the number of points. Even with a value for $\mathrm{DM}$ fluctuations higher than the values assumed in Ref. \cite{Lemos2023}, our results present uncertainties smaller than those in Ref. \cite{Lemos2023}, which uses a sample with 17 events. Therefore, these results show that larger samples can improve the constraints (see \cite{Lemos2023_2} for a discussion about fluctuations and larger samples of FRBs). 

Finally, it is worth mentioning that although DES, DESI, and Pantheon datasets point out a discrepancy in the estimate of the cosmological parameters, especially those of dark energy, here the constraints of the $f_{\mathrm{IGM}}$ are in concordance for these datasets. These results show the power of our method and the data.

\begin{table*}[h]
\vspace{0.3cm}
\centering
\caption{Estimates of the evidence and Bayes' factor.}
\begin{tabular}{ c  c  c }
\hline
Dataset &  $\ln \mathcal{E}_{i}$ & $\ln{B_{ij}}$\\
\hline
\textbf{Constant case}    &  & \\
FRB + DES Y5  &  - 96.181 $\pm$ 0.002 & -\\
FRB + DESI DR2  &  - 95.183 $\pm$ 0.002 & -\\
FRB + Pantheon &  - 95.705 $\pm$ 0.002 & -\\
\hline
\hline
\textbf{Time-dependent case}    &  & \\
FRB + DES Y5  &  - 96.712 $\pm$ 0.003 & - 0.531 $\pm$ 0.004\\
FRB + DESI DR2  &  - 95.896 $\pm$ 0.002 & - 0.713 $\pm$ 0.003\\
FRB + Pantheon &  - 95.366 $\pm$ 0.002 & + 0.339 $\pm$ 0.003\\
\hline
\end{tabular}
\label{tab:results2}
\end{table*}

One interesting method to evaluate the performance of these two models studied and compare which of them the data prefers is the Bayesian model comparison, more precisely, from the Bayes' factor calculation. Such analysis can assess if the extra complexity of a given model or parameterization (here represented by the parameter $\alpha$ present in the time-dependent case) is required by the data, preferring the model that describes the data well over a large fraction of their prior volume (see e.g.~\cite{Trotta2008,Trotta2017} for a detailed discussion). The Bayes’ factor $B_{ij}$ can be calculated from relation

\begin{equation}
    B_{ij} = \frac{\mathcal{E}_{i}}{\mathcal{E}_{j}}\;,
\vspace{0.2cm}
\end{equation}
defining the evidence as the marginal likelihood of the models, $\mathcal{E}_{i}$ and $\mathcal{E}_{j}$ correspond to the evidence of parameterizations $\mathcal{P}_{i}$ and $\mathcal{P}_{j}$, respectively. The Jeffreys’ scale \cite{Jeffreys} can be adopted to interpret the values of $\ln{B_{ij}}$ for the reference parameterization $\mathcal{P}_{j}$, where
$\ln{B_{ij}} = 0 - 1$, $\ln{B_{ij}} = 1 - 2.5$, $\ln{B_{ij}} = 2.5 - 5$, and $\ln{B_{ij}} > 5$ indicate, respectively, an inconclusive, weak, moderate and strong preference of the parameterization $\mathcal{P}_{i}$ with respect to $\mathcal{P}_{j}$. While negative values of $\ln{B_{ij}}$ mean preference in favour of $\mathcal{P}_{j}$.

Using the MultiNest algorithm \cite{MultiNest1,MultiNest2,MultiNest3} to compute the Bayesian evidence ($\ln \mathcal{E}$) and adopting the constant case as reference, we calculate the Bayes' factor. In Table \ref{tab:results2}, we present the evidence and the Bayes' factor ($\ln B_{ij}$) for each dataset. Since our results for the Bayes' factor are negative for FRB + DES Y5 and FRB +  DESI DR2 datasets, they indicate weak evidence in favor of constant parameterization. For FRB + Pantheon+, the result indicates weak evidence in favor of the time-dependent case. Therefore, these results show that a conclusive answer about the time evolution of $f_{\mathrm{IGM}}$ cannot be achieved from the current FRB observational data.

\section{Conclusions and discussions}
\label{sec:conclusions}

The latest BAO observation from DESI DR2 reports a possible tension in the $\Lambda$CDM model, indicating dynamical dark energy over a cosmological constant. This result motivates the exploration of alternative combinations of independent datasets to estimate the cosmological parameters. 

In our previous work \cite{Lemos2023}, we present a cosmological-model independent method, where we write $\mathrm{DM}_{\mathrm{IGM}}(z)$ expression in terms of $d_{L} (z)$, to constrain the $f_{\mathrm{IGM}}$. We combined 17 FRBs with SNe from the Pantheon catalog and assumed two behaviors for $f_{\mathrm{IGM}}$: the constant and time-dependent parameterizations. In the present paper, we revisited this paper, expanding the dataset by combining 107 well-localized FRBs with SNe measurements from Pantheon+ and DES Y5 and BAO data from DESI DR2. 

We estimate $f_{\mathrm{IGM,0}} \sim 0.99^{+0.01}_{-0.07}$ ($1\sigma$ level) for the constant case and $f_{\mathrm{IGM,0}}$ ranging from $0.935_{-0.211}^{+0.020}$ (FRB + DESI DR2) to $0.972_{-0.225}^{+0.009}$ (FRB + Pantheon+) at $1\sigma$ for time-dependent case. These results are consistent with our previous paper \cite{Lemos2023} (except for the time-dependent case where $\delta = 0$ and $10$ pc/cm$^{3}$) and other works in the literature that used the same parameterization \cite{Wei2019,Dai2021}. To evaluate the performance of these two models, we also performed a Bayesian model comparison. The results showed weak evidence in favor of constant parameterization (FRB + DES and FRB + DESI DR2) and weak evidence\footnote{We also performed the analysis adopting $H_{0}$ from the Planck collaboration \cite{Planck2018}, $H_{0} = 67.36 \pm 0.54$, along with a value of MB $M_{B}$ consistent with this $H_{0}$. The posterior distributions of the parameters remain similar, and the resulting Bayes factor continues to indicate only weak evidence in favor of the constant case across all datasets.} in favor of the time-dependent case (FRB + Pantheon+). 

Finally, we can note that the results indicate a concordance for the datasets used, even though they estimate cosmological parameters in disagreement, showing the robustness of our method. Also, the results present an improvement compared to the previous work, which could be attributed to the significantly different sample sizes used in these studies, reinforcing the interest in searching for a larger sample of FRBs.



\section*{Acknowledgements}

TL thanks the financial support from the Conselho Nacional de Desenvolvimento Cient\'{\i}fico e Tecnol\'ogico (CNPq). This work was developed thanks to the High-Performance Computing Center at the National Observatory (CPDON).



\newpage
\begin{longtable}[c]{l c c c c l}
\caption{Properties of FRBs}\\
\hline
\hline
\hline
Name & $z$ & $\mathrm{DM}_{\mathrm{ISM}}$ & $\mathrm{DM}_{\mathrm{obs}}$ & $\sigma_{\mathrm{obs}}$ & Refs. \\
& & [pc/cm$^{3}$] & [pc/cm$^{3}$] & [pc/cm$^{3}$] & \\
\hline
\endfirsthead

\hline
\hline
\hline
Name & $z$ & $\mathrm{DM}_{\mathrm{ISM}}$ & $\mathrm{DM}_{\mathrm{obs}}$ & $\sigma_{\mathrm{obs}}$ & Refs. \\
& & [pc/cm$^{3}$] & [pc/cm$^{3}$] & [pc/cm$^{3}$] & \\
\hline
\endhead

\hline
\endfoot

\hline
FRB 20121102A	&	0.19273	&	188.0	&	557.0	&	2.0	&	\cite{FRB121102}	\\
FRB 20180301A	&	0.3305	&	152.0	&	536.0	&	8.0	&	\cite{FRB20191228}	\\
FRB 20180814	&	0.068	&	87.75	&	189.4	&	0.4	&	\cite{FRB180814}	\\
FRB 20180916B	&	0.0337 	&	200.0	&	348.80	&	0.2	&	\cite{FRB180916}	\\
FRB 20180924B	&	0.3214	&	40.5	&	361.42	&	0.06	&	\cite{FRB180924}	\\
FRB 20181112A	&	0.4755	&	102.0	&	589.27	&	0.03	&	\cite{FRB181112}	\\
FRB 20181220A	&	0.2746	&	122.81	&	208.66	&	1.62	&	\cite{FRB181220A}	\\
FRB 20181223C	&	0.03024	&	19.9	&	112.45	&	0.01	&	\cite{FRB181220A}	\\
FRB 20190102C	&	0.2913	&	57.3	&	363.6	&	0.3	&	\cite{FRB190102}	\\
FRB 20190110C	&	0.12244	&	37	&	221.92	&	0.01	&	\cite{FRB190110C}	\\
FRB 20190303A	&	0.064	&	29.39	&	222.4	&	0.7	&	\cite{FRB180814}	\\
FRB 20190418A	&	0.07132	&	70.2	&	182.78	&	1.62	&	\cite{FRB181220A}	\\
FRB 20190425A	&	0.03122	&	49.25	&	127.78	&	1.62	&	\cite{FRB181220A}	\\
FRB 20190523A	&	0.66	&	37.0	&	760.8	&	0.6	&	\cite{FRB190523_1,FRB190523_2}	\\
FRB 20190608B	&	0.1178	&	37.2	&	338.7	&	0.5	&	\cite{FRB190608}	\\
FRB 20190611B	&	0.378	&	57.83	&	321.4	&	0.2	&	\cite{FRB190523_2}	\\
FRB 20190711A	&	0.522	&	56.4	&	593.1	&	0.4	&	\cite{FRB190523_2}	\\
FRB 20190714A	&	0.2365	&	38.0	&	504.13	&	2.0	&	\cite{FRB190523_2}	\\
FRB 20191001A	&	0.234	&	44.7	&	506.92	&	0.04	&	\cite{FRB190523_2}	\\
FRB 20191106C	&	0.10775	&	25	&	333.40	&	0.2	&	\cite{FRB190110C}	\\
FRB 20191228A	&	0.2432	&	33.0	&	297.5	&	0.05	&	\cite{FRB20191228}	\\
FRB 20200223B	&	0.06024	&	46	&	202.268	&	0.007	&	\cite{FRB190110C}	\\
FRB 20200430A	&	0.16	&	27.0	&	380.25	&	0.5	&	\cite{FRB190523_2}	\\
FRB 20200906A	&	0.3688	&	36	&	577.8	&	0.02	&	\cite{FRB20191228}	\\
FRB 20201123A	&	0.0507	&	251.93	&	433.55	&	0.0036	&	\cite{FRB201123A}	\\
FRB 20201124A	&	0.098	&	123.2	&	413.52	&	0.5	&	\cite{FRB201124}	\\
FRB 20210117A	&	0.2145	&	34.4	&	730.0	&	1.0	&	\cite{FRB210117A}	\\
FRB 20210320	&	0.2797	&	42.2	&	384.8	&	0.3	&	\cite{FRB210320}	\\
FRB 20210410D	&	0.1415	&	56.2	&	578.78	&	2.0	&	\cite{FRB210410D}	\\
FRB 20210603A 	&	0.1772	&	40.0	&	500.147	&	0.004	&	\cite{FRB210603A}	\\
FRB 20210807D	&	0.12927	&	121.2	&	251.9	&	0.2	&	\cite{FRB210807D}	\\
FRB 20211127I	&	0.0469	&	42.5	&	234.83	&	0.08	&	\cite{FRB210807D}	\\
FRB 20211203C 	&	0.3439	&	63.4	&	636.2	&	0.4	&	\cite{FRB211203C}	\\
FRB 20211212A	&	0.0715	&	27.1	&	206.0	&	5.0	&	\cite{FRB210807D}	\\
FRB 20220105A	&	0.2785	&	22.0	&	583	&	1.0	&	\cite{FRB211203C}	\\
FRB 20220204A	&	0.4	&	50.7	&	612.2	&	0.05	&	\cite{FRB220204A}	\\
FRB 20220207C	&	0.043040	&	79.3 	&	262.38	&	0.01	&	\cite{FRB220207C}	\\
FRB 20220208A	&	0.351	&	101.6	&	437	&	0.6	&	\cite{FRB220204A}	\\
FRB 20220307B	&	0.248123	&	135.7 	&	499.27 	&	0.06	&	\cite{FRB220207C}	\\
FRB 20220310F	&	0.477958	&	45.4 	&	462.24	&	0.005	&	\cite{FRB220207C}	\\
FRB 20220330D 	&	0.3714	&	38.6	&	468.1	&	0.85	&	\cite{FRB220204A}	\\
FRB 20220418A	&	0.622000	&	37.6 	&	623.25	&	0.01	&	\cite{FRB220207C}	\\
FRB 20220501C	&	0.381	&	31	&	449.5	&	0.2	&	\cite{FRB210807D}	\\
FRB 20220506D	&	0.30039 	&	89.1	&	396.97	&	0.02	&	\cite{FRB220207C}	\\
FRB 20220509G	&	0.089400	&	55.2	&	269.53	&	0.02	&	\cite{FRB220207C}	\\
FRB 20220610A	&	1.016	&	31.0	&	1458.15	&	0.2	&	\cite{FRB20220610A}	\\
FRB 20220717A	&	0.36295	&	118	&	637.34	&	3.52	&	\cite{FRB220717A}	\\
FRB 20220725A	&	0.1926	&	31	&	290.4	&	0.3	&	\cite{FRB210807D}	\\
FRB 20220726A	&	0.361	&	89.5	&	686.55	&	0.01	&	\cite{FRB220204A}	\\
FRB 20220825A	&	0.241397	&	79.7	&	651.24	&	0.06	&	\cite{FRB220207C}	\\
FRB 20220831A 	&	0.262	&	1019.50	&	1146.25	&	0.2	&	\cite{FRB220204A}	\\
FRB 20220912A	&	0.0771	&	125.00	&	219.46	&	0.042	&	\cite{FRB220912A}	\\
FRB 20220914A	&	0.113900 	&	55.2	&	631.28	&	0.04	&	\cite{FRB220207C}	\\
FRB 20220918A	&	0.491	&	41	&	656.8	&	0.8	&	\cite{FRB210807D}	\\
FRB 20220920A	&	0.158239	&	40.3	&	314.99 	&	0.01	&	\cite{FRB220207C}	\\
FRB 20221012A	&	0.284669	&	54.4	&	441.08	&	0.7	&	\cite{FRB220207C}	\\
FRB 20221106A	&	0.2044	&	35	&	343.8	&	0.8	&	\cite{FRB210807D}	\\
FRB 20221219A 	&	0.554	&	44.4	&	706.7	&	0.6	&	\cite{FRB221219A}	\\
FRB 20230526A	&	0.1570	&	50	&	361.4	&	0.2	&	\cite{FRB210807D}	\\
FRB 20230708A	&	0.105	&	50	&	411.51	&	0.05	&	\cite{FRB210807D}	\\
FRB 20230718A	&	0.035	&	396	&	477.0	&	0.5	&	\cite{FRB210807D}	\\
FRB 20230814A	&	0.5535	&	104.9	&	696.35	&	0.5	&	\cite{FRB220204A}	\\
FRB 20230902A	&	0.3619	&	34	&	440.1	&	0.1	&	\cite{FRB210807D}	\\
FRB 20231226A	&	0.1569	&	145	&	329.9	&	0.1	&	\cite{FRB210807D}	\\
FRB 20240114A	&	0.13	&	49.7	&	527.65	&	0.01	&	\cite{FRB240114A}	\\
FRB 20240201A	&	0.042729	&	38	&	374.5	&	0.2	&	\cite{FRB210807D}	\\
FRB 20240209A	&	0.1384	&	55.5	&	176.49	&	0.01	&	\cite{FRB240209A}	\\
FRB 20240210A	&	0.023686	&	31	&	283.73	&	0.05	&	\cite{FRB210807D}	\\
FRB 20240310A	&	0.1270	&	36	&	601.8	&	0.2	&	\cite{FRB210807D}	\\
\label{tab:data}
\end{longtable}
\end{document}